\documentclass{article}
\usepackage{graphicx}

\begin{document}

\title
{Measurement of qutrits}

\author{A.V. Burlakov, L.A. Krivitskiy, S.P. Kulik, G.A. Maslennikov, M.V. Chekhova}

\date{}
\maketitle
\begin{center}
{Quantum Electronics Division, Physics Department, Moscow State
University, 119899 Moscow, Russia.
e-mail:postmast@qopt.phys.msu.su}
\end{center}

\begin{abstract}
We proposed the procedure of measuring the unknown state of the
three-level system -- the qutrit, which was realized as the
arbitrary polarization state of the single-mode biphoton field.
This procedure is accomplished for the set of the pure states of
qutrits; this set is defined by the properties of SU(2)
transformations, that are done by the polarization transformers.
\end{abstract}

\section{Introduction}

In the physics of quantum information those systems that can be completely
described in terms of three orthogonal states, were called qutrits (q-trit).
In the case of a pure state the wave function of the three-level system can
be written as:

\begin{equation}
\label{eq1}
\Psi = c_1 \left| 1 \right\rangle + c_2 \left| 2 \right\rangle + c_3 \left|
3 \right\rangle ,
\end{equation}

\noindent
where $|$1$\rangle$, $|$2$\rangle$, $|$3$\rangle$ - are
the orthogonal basis states. The complex coefficients $c_j $ are
the amplitudes of the basis states $\left| j \right\rangle $ and
satisfy the following normalizing condition

\begin{equation}
\label{eq2}
\sum\limits_{i = 1}^3 {\left| {c_j } \right|^2} = 1.
\end{equation}

Decomposition (1) is the generalization of the definition of
qubit when the dimension of the quantum system $d = 3$.

Several ways are known how to experimentally realize the
multilevel quantum optical systems. In one of them [1] the
interferometric procedure of state preparation is used, when the
attenuated laser pulses are sent into the multi-armed
interferometer. The number of arms is equal to the system's
dimensionality. Identification of basis states is done either
through the pulses delay (temporal basis) or by the presence of
constructive interference in the certain arm of the
interferometer that is put into the registration system (energy
basis). The other example is the optical field that consists of
the pairs of correlated photons that belong to the different
polarization modes. The preparation of such fields and their
unitary transformations are reviewed in [2,3]. Multi-level systems
attract a great interest in quantum cryptography, where by using
those systems one can achieve secrecy growing when the
eavesdropper uses so called symmetric individual attacks [4-6].

The problem of the adequate measurement of the parameters of the
quantum state is one of the major ones in quantum information
science. With the decision of this problem one can expect the
realization of the information's output devices, the protocols of
error correction, quantum repeaters, and other quantum
communication devices. From the fundamental point of view the
question of minimal set of measurements that is needed for the
complete description of the system's state is also very
important. Let us notice that in some cases it is not necessary
to perform a complete set of measurements to define system's
purity [7].

Of course, for different types of quantum states one also uses
different types of measurement procedures. As an example for a
squeezed state of light, methods of homodyne tomography are
developing [8]; they, in principle allow restoring the density
matrix of n-photon Fock states [9]. For the polarization-squeezed
[10] and scalar [11] light the fluctuations of Stokes parameters
are analyzed and the quasi-probability function is restored [12].
In the case of two-photon fields one registers the set of the
fourth order field moments in different spatial and polarization
modes [13]. Let us notice that in context of each experimental
procedure, the \textit{a-priori} information about the properties
of the examined state plays an important role.

\section{Biphotons as qutrits}

This work is devoted to the optical realization of the protocol
that allows restoring the density matrix of an unknown
three-level system quantum state. The polarization state of the
two-photon field that belongs to the single spatial and frequency
mode is the object of our investigation. In [2] it was shown that
the pure state of such a field can be written as:

\begin{equation}
\label{eq3} \Psi = c_1 \left| {2,0} \right\rangle + c_2 \left|
{1,1} \right\rangle + c_3 \left| {0,2} \right\rangle .
\end{equation}

The two-photon Fock states in two orthogonal; polarization modes
H and V serve as the basis states. So, for example, the second
term in (3) corresponds to the existence of one photon in modes H
and V with $\left| {c_2 } \right|^2$ probability. The vacuum
component $\left| {0,0} \right\rangle $ is not considered in (3),
because when the field is registered by the method of coincidence
of photopulses, the contribution of this component to the
measured correlation functions is equal to zero. The imaginary
parts of the complex coefficients $c_j = \left| {c_j }
\right|\exp \left\{ {i\varphi _j } \right\}, j = 1,2,3$ are the
phases of the basis states. The total phase of the wave function
is incidental, and that's why one of the phases can be excluded
from the consideration, giving us the relative phases. For
example, $\varphi _{12} = \varphi _1 - \varphi _2 $ and $\varphi
_{32} = \varphi _3 - \varphi _2 $.

For the description of the polarization properties of the
single-mode biphoton field, in [14] the so-called polarization or
fourth-order coherency matrix was introduced:

\begin{equation}
\label{eq4}
K_4 = \left( {{\begin{array}{*{20}c}
 A \hfill & D \hfill & E \hfill \\
 {D^\ast } \hfill & C \hfill & F \hfill \\
 {E^\ast } \hfill & {F^\ast } \hfill & B \hfill \\

\end{array} }} \right).
\end{equation}

The elements of this matrix represent the normally ordered fourth moments of
the field that can be written as

\[
\begin{array}{l}
 A \equiv \left\langle {a^{\dagger2}a^{2}}\right\rangle, \quad B \equiv
\left\langle {b^{\dagger2}b^{2}}\right\rangle, \quad
C\equiv\left\langle {a^{\dagger}b^{\dagger}ab}\right\rangle, \\
 D \equiv \left\langle {a^{\dagger2}ab}\right\rangle, \quad E \equiv
\left\langle {a^{\dagger2}b^{2}} \right\rangle, \quad F \equiv
\left\langle {a^{\dagger}b^{\dagger}b^{2}}\right\rangle

. \\
 \end{array}
\]

Here, $a^\dag \equiv a_H^\dag$, $b^\dag \equiv a_V^\dag$, $a
\equiv a_H$, $b \equiv a_V$
 - are the operators of photon
creation and annihilation in polarization modes $H$ and $V$,
respectively. It can be noticed that the diagonal components of
$K_4 $ are real. They characterize the intensity fluctuations in
parallel ($A$ and $B$) or orthogonal ($C$) polarization modes.
Non-diagonal elements $D$, $E$, $F$ in general case are complex.

Since the state of biphoton field can be fully described by the
fourth moments of the field, the elements of $K_4 $ matrix can be
derived through the components of the density matrix of biphoton
field. As an example for the pure state (3), by definition [15]
$\rho _{mn} = c_m c_n^\ast $ and

\begin{equation}
\label{eq5} \rho _{11} = \left| {c_1 } \right|^2 = A/2,
\quad\rho_{22} = \left| {c_2 } \right|^2 = C, \quad \rho _{33} =
\left| {c_3 } \right|^2 = B/2 ,
\end{equation}

\begin{equation}
\label{eq6} \rho _{12} = c_1 c_2^\ast = {D^\ast } / {\sqrt 2 },
\quad \rho _{13} = c_1 c_3^\ast = {E^\ast } / 2, \quad \rho _{23}
= c_2 c_3^\ast = {F^\ast } / {\sqrt 2 }.
\end{equation}

\noindent
 The condition of state's purity

\begin{equation}
\label{eq7}
\rho ^2 = \rho .
\end{equation}

\noindent
And normalizing condition

\begin{equation}
\label{eq8}
Sp\left( \rho \right) = 1
\end{equation}

\noindent impose the certain constraints between the elements of
$K_4$ matrix. Thus, from (8) it follows that

\begin{equation}
\label{eq9}
A + B + 2C = 2,
\end{equation}

\noindent And condition (7) gives

\begin{equation}
\label{eq10} E^\ast = {ABC} \mathord{\left/ {\vphantom {{ABC}
{DF}}} \right. \kern-\nulldelimiterspace} {DF}, \quad \left| F
\right|^2 = BC\mbox, \quad \left| D \right|^2 = C(2 - B - 2C).
\end{equation}

For the mixed state, the definition of density matrix also
includes the complementary averaging with the classical
distribution function $P$ by the possible states of the system,
where $P$ satisfies $\sum\limits_{i=1}{P_i }= 1$, and the
components of density matrix are:

\begin{equation}
\label{eq11}
\rho _{mn} = \overline {c_m c_n^\ast } .
\end{equation}

\section{Measurement of the state of qutrits}

The question rises -- how many (real) parameters should be
measured to characterize completely the unknown state of biphoton
field? From the definition and properties of the density matrix,
it follows that in the case of the pure state, the number of real
parameters, that define the state of the system that has a
dimension $d$, is equal to $2d - 2$, and in the case of a mixed
state is equal to $d^2 - 1$. Correspondingly for the qutrits in
the first case one needs to know four real numbers, in the second
case - eight. Considering that in experiment one measures the
unnormalized state's amplitudes and conditions (8,9) are needed
to be checked every time after measuring all three diagonal
elements of $K_4 $ matrix, we obtain that in pure state five
moments are needed to be measured and in the mixed state - nine.

Before we go further into the discussion of suggested protocol, we notice
that the measurement procedure always leads to the destruction of our state
that is caused by its interaction with the classical measuring device, in
our case -- with the detector. So when we speak about the input state, we
always have in mind that it is introduced by a large enough set of copies,
and part of them can be destroyed by the measurement. The results of the
measurements will be applied to the rest part of the ensemble; this
procedure lies at the heart of the ensemble method of quantum measurements.

In quantum optics as the measuring apparatus for the fourth
moments of the field usually serves the Brown-Twiss scheme, which
consists of the beam splitter with the photo detectors in the
output ports (Fig.1). Polarization transformations in each
spatial mode is done with the help of retardation plates
($\lambda $/2 è $\lambda $/4) and polarization filters
(polarizers) Ð. Let's examine the normally ordered fourth moments
of the field $R_{12} \left( {\chi _1 ,\theta _1 ,\chi _2 ,\theta
_2 } \right ) \sim \left\langle {{b'}_1^\dag {b'}_2^\dag {b'}_1
{b'}_2} \right\rangle $, that are registered in the scheme that
is shown at Fig.1, with the given wave plates orientation and the
fixed polarization that is transmitted by polarizers. Our goal
will be \textit{the search of the minimal set} of such moments
i.e. parameters that are measured in the experiment, from which
one can compose all the elements of $K_4 $ matrix. In this case
the input field will be transformed by the wave plates and
polarizers in such way that the registered moments can be derived
through the components of $K_4 $ matrix. Polarization
transformations that are done by the wave plates are unitary and
the polarizer plays the role of the polarization filter that sets
the polarization state registered by the detectors. This idea is
based on the classical schemes in which the Stokes parameters are
measured [16]. It was also used in [13] for the measurement of
the polarization properties of the biphoton light in two spatial
modes (so-called two-qubit case). Let's notice that the choice of
quarter- and half- wave plates as the polarization transformers
evidently is not unique, it is dictated by the convenience (these
wave plates are most commonly used in the polarization
experiments) and by the clearance of the transformations.

The consecutive action of the beam splitter, two wave plates and
the polaraizer that transmits the vertical polarization on the
signal (idler) photon is described by the following matrix
transformations:

\begin{equation}
\label{eq12} \left( {{\begin{array}{*{20}c}
 {a}' \hfill \\
 {b}' \hfill \\
\end{array} }} \right) = G_V G_{\lambda/2} G_{\lambda/4} G_{BS} \left(
{{\begin{array}{*{20}c}
 a \hfill \\
 b \hfill \\
\end{array} }} \right),
\end{equation}

\noindent where $a$ and $b$ -- are the annihilation operators of
the input state in two orthogonal polarization modes $H$ and $V$
at the input, ${a}'$ and ${b}'$ - are the annihilation operators
at the output of the transformers; the state vector is written in
Jones representation.

\begin{equation} \label{eq13} G_{BS} =
\left( {{\begin{array}{*{20}c}
 {1 \mathord{\left/ {\vphantom {1 {\sqrt 2 }}} \right.
\kern-\nulldelimiterspace} {\sqrt 2 }} \hfill & 0 \hfill \\
 0 \hfill & {1 \mathord{\left/ {\vphantom {1 {\sqrt 2 }}} \right.
\kern-\nulldelimiterspace} {\sqrt 2 }} \hfill \\
\end{array} }} \right)
\end{equation}

\noindent
is the matrix that describes the action of the non-polarizing beam splitter,

\begin{equation}
\label{eq14} G_V = \left( {{\begin{array}{*{20}c}
 0 \hfill & 0 \hfill \\
 0 \hfill & 1 \hfill \\
\end{array} }} \right)
\end{equation}

\noindent
is the matrix of the polarizer that transmits the vertical field component,

\begin{equation}
\label{eq15} G_{\lambda/4,\lambda/2}= \left(
{{\begin{array}{*{20}c}
 t \hfill & r \hfill \\
 { - r^\ast } \hfill & {t^\ast } \hfill \\
\end{array} }} \right)
\end{equation}

\noindent
are the matrices of the wave plates. Here coefficients $t$ and $r$ are equal to

\begin{equation}
\label{eq16}
\begin{array}{l}
 t = \cos \delta + i\sin \delta \cos 2\alpha , \\
 r = i\sin \delta \sin 2\alpha , \\
 \end{array}
\end{equation}

\noindent where $\delta $ - is the optical thickness, and $\alpha
$ - is the angle between optical axis and the vertical direction
($V)$. For the quarter and half- wave plates $\delta _{\lambda
/4} = \pi/4 , \quad\alpha _{\lambda /4} \equiv \chi , \quad
\delta _{\lambda /2}  = \pi/ 2 \, \quad\alpha _{\lambda /2}
\equiv \theta ,$ and these coefficients can be rewritten as:

\begin{equation}
\label{eq17}
\begin{array}{l}
 t_{\lambda /4} = 1 \mathord{\left/ {\vphantom {1 {\sqrt 2 }}}
\right. \kern-\nulldelimiterspace} {\sqrt 2 }(1 + i\cos 2\chi ), \\
 r_{\lambda /4} = i \mathord{\left/ {\vphantom {i {\sqrt 2 }}}
\right. \kern-\nulldelimiterspace} {\sqrt 2 }\sin 2\chi , \\
 \end{array}
\end{equation}

\begin{equation}
\label{eq18}
\begin{array}{l}
 t_{\lambda /2} = i\cos 2\theta , \\
 r_{\lambda /2} = i\sin 2\theta . \\
 \end{array}
\end{equation}

It is clearly seen from the definition of $K_4 $ matrix how one
can measure its diagonal components or moments $A, B$ and $C$. In
the first case the optical axes of all wave plates are set
vertically - along the direction of the transmission of polarizer
$P$. In the second case the polarization in two arms must be
rotated by 90$^{0}$, that is achieved by setting $\chi _1 = 0^o,
\theta _1 = 45^o$, $\chi _2 = 0^o,\theta _2 = 45^o$. In the third
case the polarization is rotated only in one arm, and according to
scheme's symmetry it is not important in which. The set is: $\chi
_1 = 0^o, \theta _1 = 45^o$, $\chi _2 = 0^o,\theta _2 = 0^o$. The
more complex transformations are needed when non-diagonal
components of $K_4 $ matrix are measured. As an example let's
look how the following setting of elements acts on the input
state does. Let $\chi _1 = 0^o, \theta _1 = 45^o$, $\chi _2 =
45^o, \theta _2 = 22.5^o$. It is not hard to calculate that in
this case

\begin{equation}
\label{eq19}
\begin{array}{l}
 R_{12} = \left\langle {{b'}_1^\dag {b'}_2^\dag {b'}_1 {b'}_2 }
\right\rangle = 1 \mathord{\left/ {\vphantom {1 8}} \right.
\kern-\nulldelimiterspace} 8\left[ {\left\langle {a^{\dag2}a^2}
\right\rangle - \left\langle {a^{\dag2}ab} \right\rangle -
\left\langle {a^\dag b^\dag a^2} \right\rangle + \left\langle
{a^\dag b^\dag ab}
\right\rangle } \right] = \\
 1 \mathord{\left/ {\vphantom {1 8}} \right. \kern-\nulldelimiterspace}
8\left[ {A + C - 2ReD} \right]. \\
 \end{array}
\end{equation}

The measured moment in this case contains the contributions from
three elements of the coherency matrix. Two of them are real
diagonal components $A$, $C$ . The third one is the real part of
the (complex) non-diagonal element $D$. This example shows that
since we are not directly measuring the phases of the states
$\varphi _{1}$, $\varphi _{2}$ or $\varphi _{3}$, but its cosine
and sine, then the number of the measurements needed is
increasing. In the real experiment, which description is shown
below, in each arm of the Brown-Twiss scheme we used simpler
configuration than the set of two rotating wave plates and the
fixed polarizer. We considered the fact that when measuring the
moments of the fourth order, the transformation that is done by
the half wave plate and the fixed polarizer is equivalent to the
action of one polarizer, which orientation is given by the angle
$\beta $. The rotation angles of half wave plate $\theta $ and
the polarizer $\beta $ are bounded by the equation:

\begin{equation}
\label{eq20} \beta = - 2\theta .
\end{equation}

In the Table 1 we show the values of the orientation angles of the
quarter wave plates ($\chi _{1,2})$ and polarizers $(\beta
_{1,2})$ in two arms versus the value of the corresponding
measured moment. This table essentially serves as the protocol of
the reconstruction of the input state of the field that is
presented by the biphoton-qutrit. It can be seen that in general
case, the number of required measurements is equal to nine. First
seven measurements realize the protocol for the pure input state.
Two additional measurements are necessary for the definition of
the cosine (sine) values of the corresponding phases. Eighth and
ninth lines of the table show how to find the real and imaginary
parts of the complex moment $Å$, which in the case of the pure
state, according to (10), is derived through the rest of the
moments. Let's notice that in our protocol for the definition of
each non-diagonal elements of $K_4$ matrix, only three moments
must be known, what the minimal number of measurement needed is
apparently.

\begin{table}
\caption{\label{t.1} The protocol of measurement of the moment
set, which form the coherency matrix.}
\begin{tabular}{|c|c|c|c|c|c|}
  \hline
    &plate $ \lambda/4 $ & Polarizer &plate $ \lambda/4$ & Polarizer &  Field \\
    & (I) & (I) & (II) & (II) & Moment \\
    & $\chi_1$, deg. & $\beta_1$, deg. & $\chi_2$, deg. & $\beta_2$, deg. &  \\\hline
  1. & 0 & 90 & 0 & 90 & A/4 \\\hline
  2. & 0 & 90 & 0 & 0 & C/4 \\\hline
  3. & 0 & 0 & 0 & 0 & B/4 \\\hline
  4. & 45 & 0 & 0 & 0 & 1/8(B+C+2ImF) \\\hline
  5. & 45 & -45 & 0 & 0 & 1/8(B+C-2ReF)  \\\hline
  6. & 45 & -45 & 0 & 90 & 1/8(A+C-2ReD)  \\\hline
  7. & 45 & 0 & 0 & 90 & 1/8(A+C+2ImD)  \\\hline
  8. & -45 & 22,5 & -45 & 22,5 & 1/16(A+C-2ImE)   \\\hline
  9. & 45 & 45 & 45 & -45 & 1/16(A+C-2ReE)   \\ \hline
\end{tabular}
\end{table}

\section{Experiment}

The experimental setup is shown at Fig.2. It can be
conventionally separated in two blocks -- the block of the
preparation of the input state and the block of measurements.
First block includes the cw Ar$^+$ laser, operating at 351 nm,
with the output power of 120 mW, which serves as the pump for the
non-linear lithium iodate crystal, where the process of biphoton
generation goes on. This block also includes the system of
adjustment mirrors, quartz wave plate, which orientation could be
smoothly varied, and the interference filter with the central
wavelength of 702 nm, and 5 nm FWHM. Two-photon states of light
were generated by the process of spontaneous parametric
down-conversion (SPDC) inside the non-linear crystal. We used
type-I collinear, frequency degenerate SPDC. The wavelength of
biphoton radiation was $\lambda _s = \lambda _i = 2\lambda _p =
702\pm 9nm$. The polarization of both photons was vertical. In
this case, right after the crystal, the biphoton field was in
state

\begin{equation}
\label{eq21} \Psi = c_3^{in} \left| {0,2} \right\rangle + \left|
{vac} \right\rangle .
\end{equation}

The quartz wave plate was used to transform this state to the one
that is described by equation (3) (Fig.2). It is known that all
the transformations that are done with the polarization of
biphotons by the retardation plates can be described by the
unitary (3$\times$3) matrices G [2]:

\begin{equation}
\label{eq22} \left( {{\begin{array}{*{20}c}
 {c_1 } \hfill \\
 {c_2 } \hfill \\
 {c_3 } \hfill \\
\end{array} }} \right)^{out} = G\left( {{\begin{array}{*{20}c}
 {c_1 } \hfill \\
 {c_2 } \hfill \\
 {c_3 } \hfill \\
\end{array} }} \right)^{in},
\end{equation}

\noindent where\noindent
\begin{equation}
G = G\left( {\delta ,\alpha } \right) =
 \left( \begin{array}{ccc}
 t^2  & \sqrt {2} tr  & r^2  \\
  - \sqrt {2} tr^\ast  & \left| t \right|^2 - \left| r \right|^2
 & \sqrt {2} t^\ast r  \\
 r^{\ast 2}  & { - \sqrt 2} t^\ast r^\ast   & t^{\ast2}
\end{array}  \right) ,
\end{equation}

\noindent and coefficients $t$ and $r$ were introduced by (16).

Matrices (23) give us the irreducible presentation of SU(2) group
with 3$\times$3 dimensionality in the space of the state vectors
(3). Let's notice that one cannot realize an arbitrary
polarization state of a biphoton field, by using wave plates
only. In general case such transformations, together with the
space of a state vectors (3) form the three-dimensional unitary
presentation of SU(3) group.

The thickness of the setting wave plate was $h $= 824$\pm $1mkm,
therefore the parameter $\delta \equiv \frac{\pi }{\lambda
}\left( {n_o - n_e } \right)h$ was fixed. The second parameter
$\alpha $ was changing during the experiment that allowed us to
set the state of a biphoton field that was given onto the input
of the measurement block. It is clear that all the states that
were prepared in such a way did not drive our biphoton field out
of the pure states class.

The measurement block consists of the Brown-Twiss scheme that is
shown at Fig.1. In our experiments we used Pockel cells instead
of the quarter wave plates. The use of the Pockel cells seemed
preferable to us, because it allowed controlling the polarization
transformation distantly, by applying the certain voltages on
them. The spectral control of the biphoton field was realized
with the help of the spectrograph. Pulses coming from the
detectors were driven onto the standard coincidence scheme, which
measured the number of coincidences rate that is proportional to
correlators (19).

The measurement procedure was as follows. For the certain
orientation of the setting wave plate, we performed a set of
measurements that is described in Table 1. Then we rotated the
setting wave plate by $\alpha $, what corresponded to the change
of the input state, and performed the same set of measurements.

The dependence of modulo squared of the three state's amplitude
and two phases from the rotation angle of a setting wave plate is
plotted on Fig.3-6. Each experimental dot on the plot corresponds
to the certain input state that is given by a setting wave plate.
Solid lines -- result of theoretical calculations by formulas
(22,23). Since all three measured moments contributed to the
phase's calculation (look at Table 1); the errors of three
measurements were added and the precision of a corresponding
measurement was low. The main source of errors is the low quality
of the Pockel cells and as a consequence -- the inadequacy of the
polarization transformations that are done with these elements.
Definitely by using the retardation wave plates that are working
in zero-order interference regime is the only way to overcome
this problem. We notion the good correspondence of the
calculations and the experimental results when measuring the
modules of state's amplitudes. All errors that appear here are
due to the errors in setting the correct polarizer orientation
angle.

The presentation of the measured field states in terms of complex
coefficients $c_{j}$  as it is done on Fig.3-6, is possible only
for the pure states. For the mixed states, one needs to measure
all six moments, that form the $K_4 $ (4) matrix. The real and
imaginary parts of the moments $D$ and $F$ are shown at Fig.7,8.
The calculated and measured components of density matrix of the
state that correspond to the orientation of the setting wave
plate by $\alpha = 25^0$, are shown at Fig. 9. Of course, for the
pure states, which case was realized in our experiment, these
moments can be derived through amplitudes $c_{i}$. Moment $Å$ was
not measured in this case, because as it is shown above, it could
be derived through the other moments.

\section{Conclusion}

In this work, we proposed the procedure of measuring the unknown
state of the three-level system -- the qutrit, which was realized
as the arbitrary polarization state of the single-mode biphoton
field. This procedure is experimentally realized for the set of
the pure states of qutrits; this set is defined by the properties
of SU(2) transformations, that are done by the polarization
transformers (retardation plates).

However, it seems to be interesting to realize the complete
protocol of density matrix restoration both for the pure and for
the mixed states of qutrits. These experiments are now in
progress and their results will be published soon. Separately,
the question of maximum-likelihood estimation of the experimental
results to the most probable quantity of $\rho $ [13,17] will be
reviewed.

This work was done by the financial support of RFBR (grant {\#} 02-02-16664)
and INTAS (01-2122).

\begin{figure}[p]

  \centering
  \includegraphics[width=1.0\textwidth]{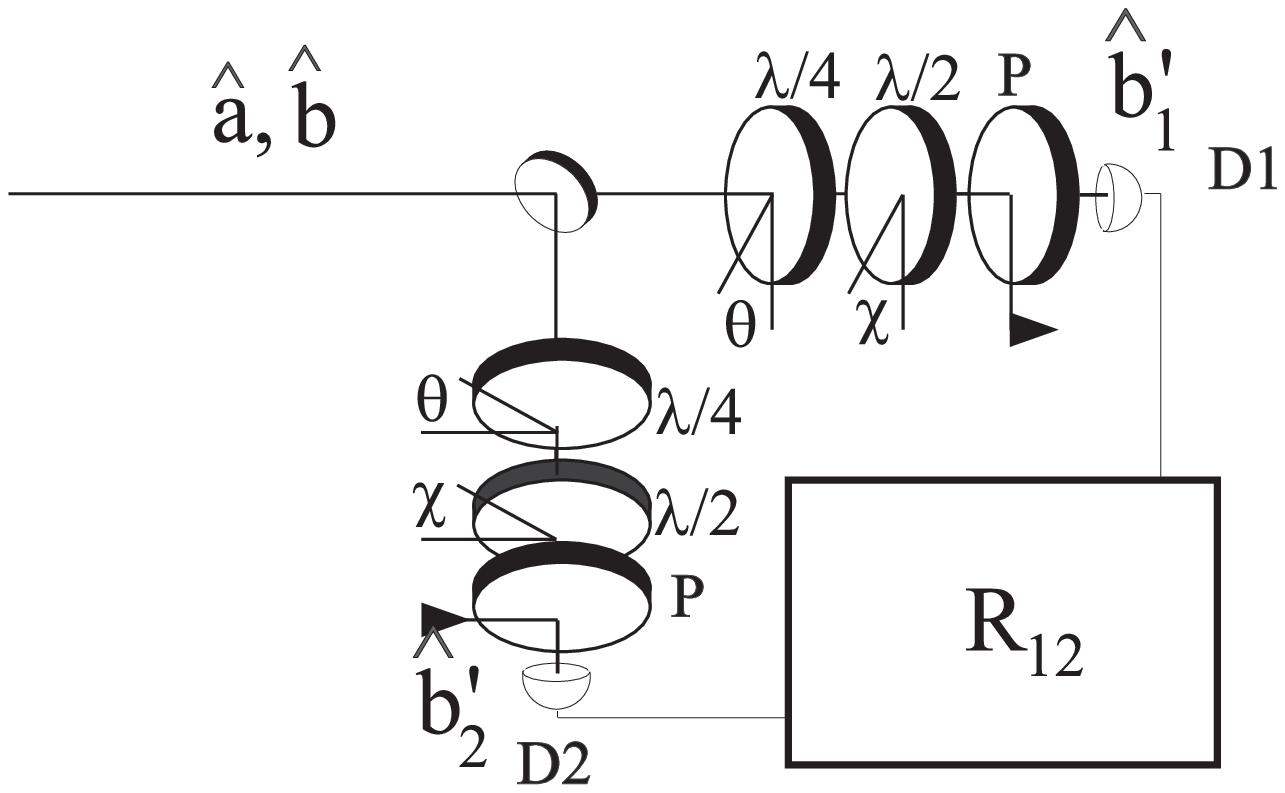}
  \caption{Brown-Twiss scheme in which one can measure the moments
that form the coherency matrix $K_4 $. BS -- beam splitter, RP --
retardation plates of quarter and half wavelength, that are
characterized by the parameters $\delta _{\lambda/2} ,\delta
_{\lambda /4}$ and $\theta _{\lambda /2 } ,\chi _{\lambda /4} $,
 P are the polarizers that transmit the vertical polarization.}
\end{figure}

\begin{figure}[p]
  \centering
  \includegraphics[width=1.0\textwidth]{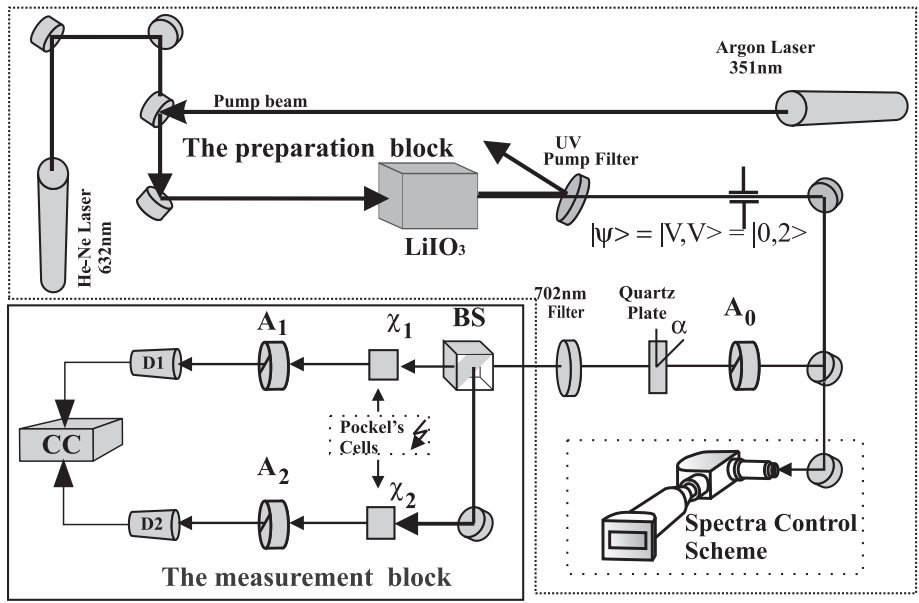}
  \caption{Experimental setup.}
\end{figure}

\begin{figure}[p]
  \centering
  \includegraphics[width=1.0\textwidth]{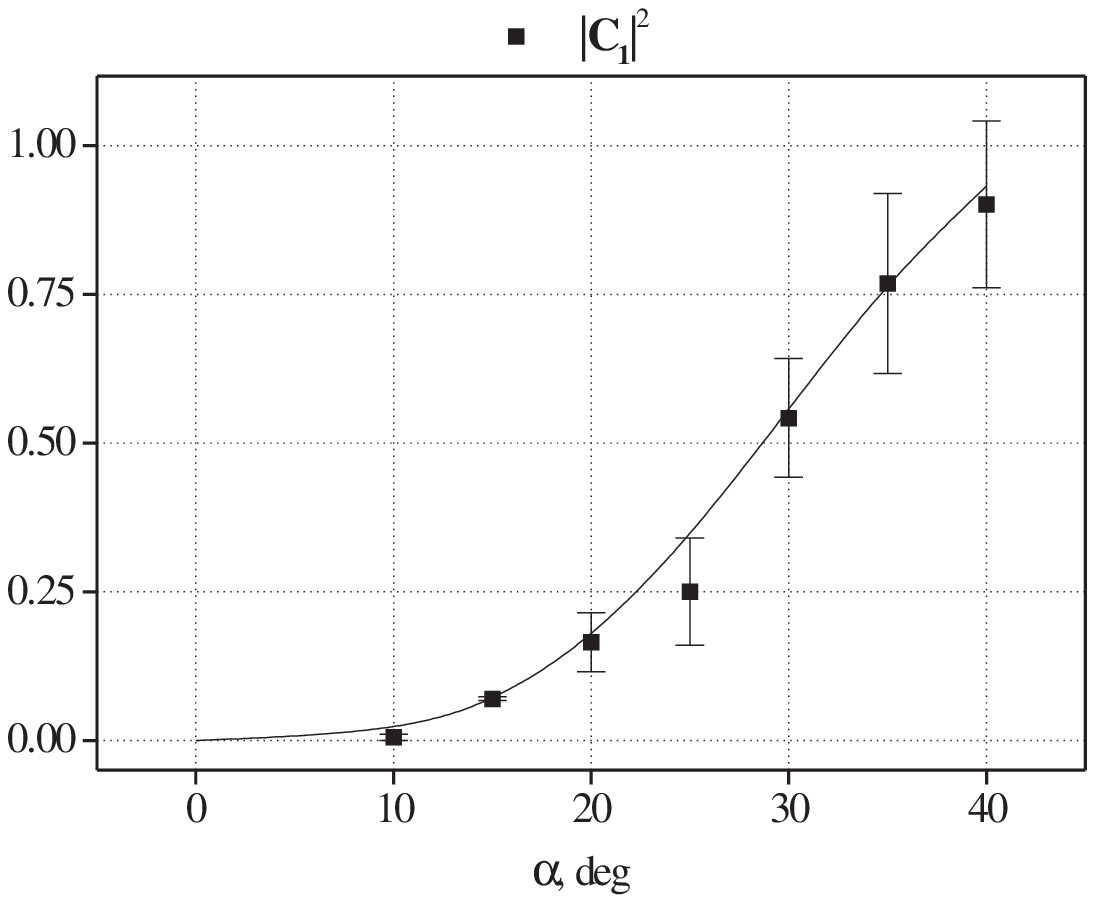}
  \caption{Dependence of the parameter of the input state $\left|
{c_1 } \right|^2$,from the rotation angle of the setting
retardation plate}
\end{figure}

\begin{figure}[p]
  \centering
  \includegraphics[width=1.0\textwidth]{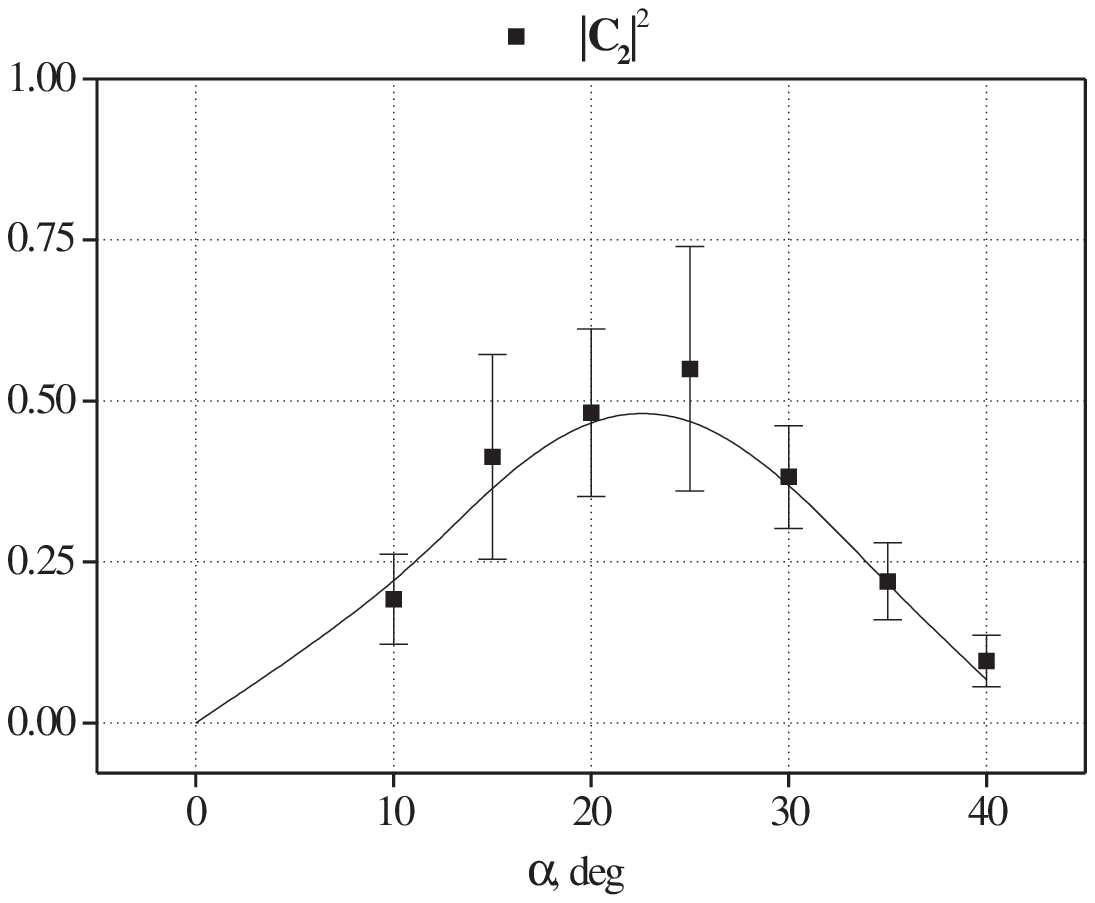}
  \caption{Dependence of the parameter of the input state $\left|
{c_2 } \right|^2$,from the rotation angle of the setting
retardation plate}
\end{figure}

\begin{figure}[p]
  \centering
  \includegraphics[width=1.0\textwidth]{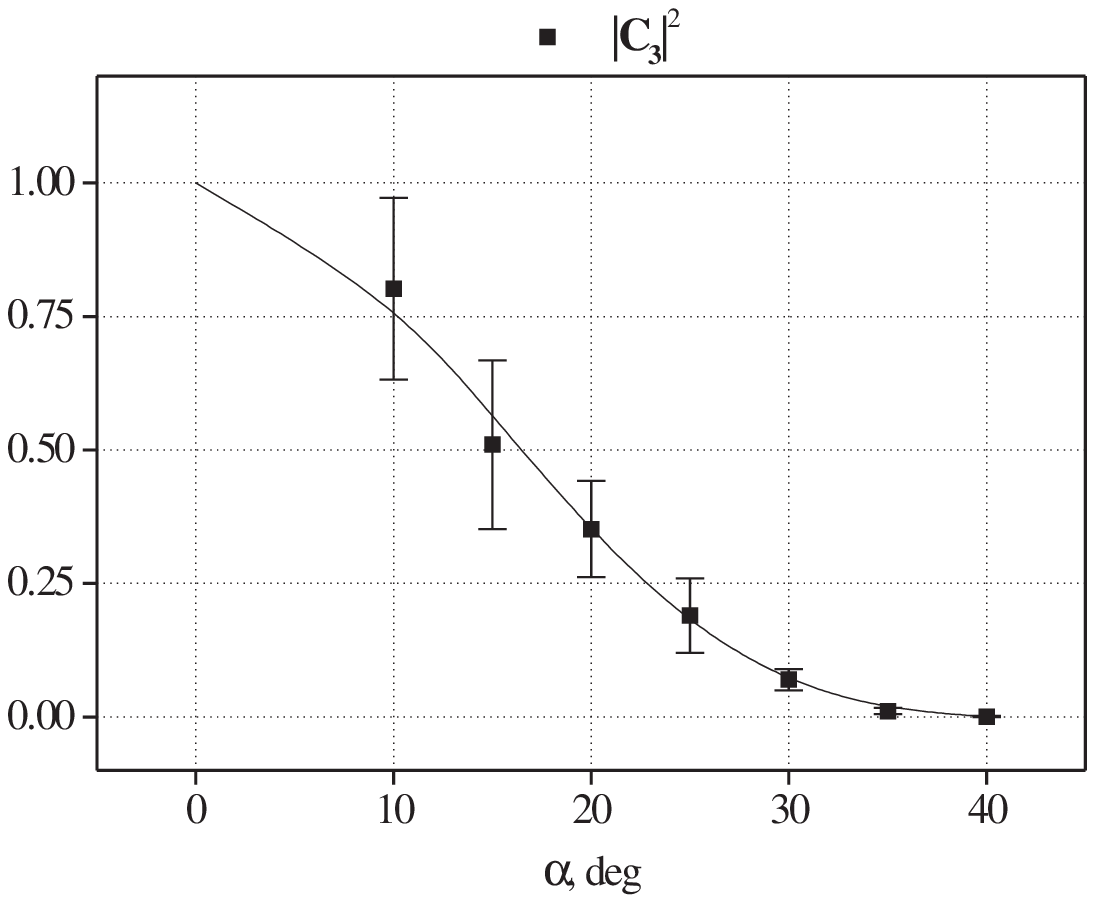}
  \caption{Dependence of the parameter of the input state $\left|
{c_3 } \right|^2$,from the rotation angle of the setting
retardation plate}
\end{figure}

\begin{figure}[p]
  \centering
  \includegraphics[width=1.0\textwidth]{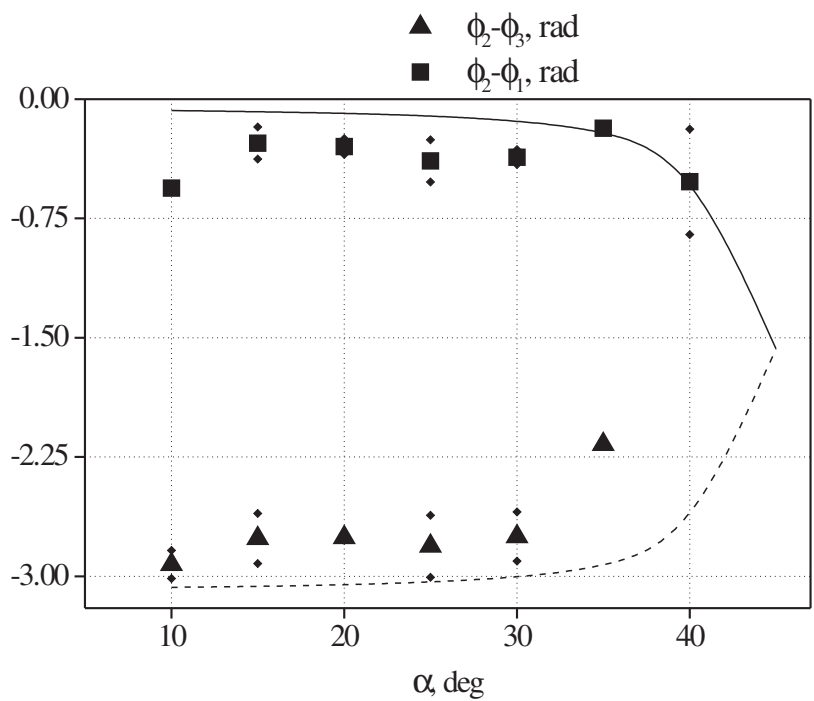}
  \caption{Dependence of the parameter of the input state $ \varphi _2 - \varphi _3 , \varphi _2 -
\varphi _1$, from the rotation angle of the setting retardation
plate}
\end{figure}

\begin{figure}[p]
  \centering
  \includegraphics[width=1.0\textwidth]{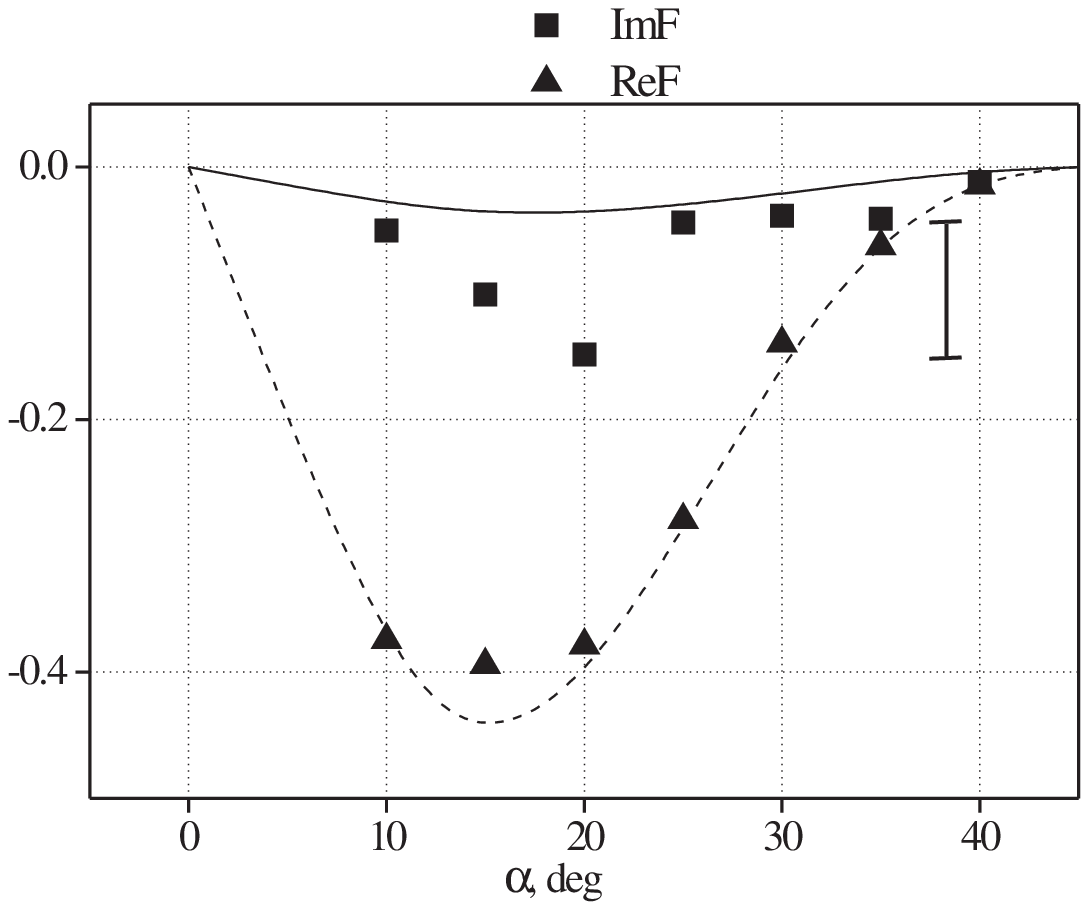}
  \caption{Dependencies of the real and imaginary parts of the moment F from the rotation
  angle of the setting retardation plate.}
\end{figure}

\begin{figure}[p]
  \centering
  \includegraphics[width=1.0\textwidth]{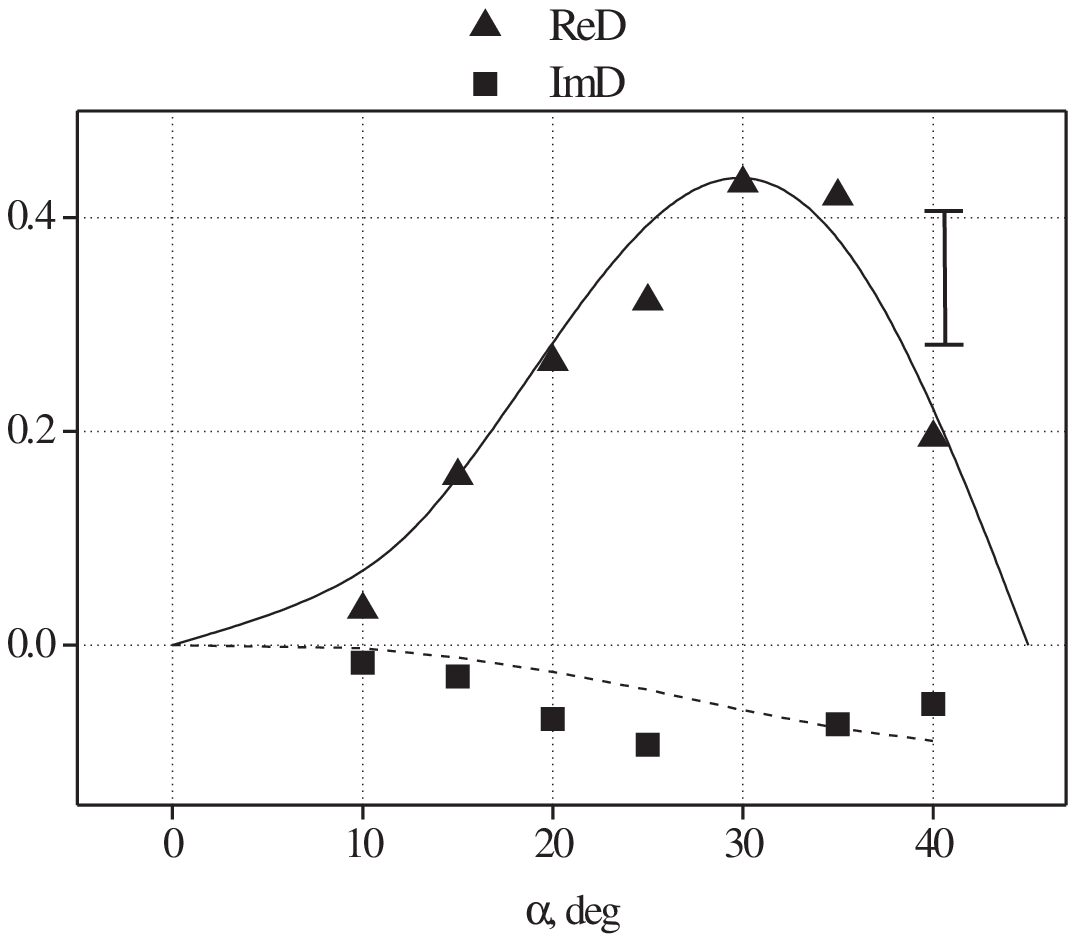}
  \caption{Dependencies of the real and imaginary parts of the moment D from the rotation
  angle of the setting retardation plate.}
\end{figure}

\begin{figure}[p]
  \centering
  \includegraphics[width=1.0\textwidth]{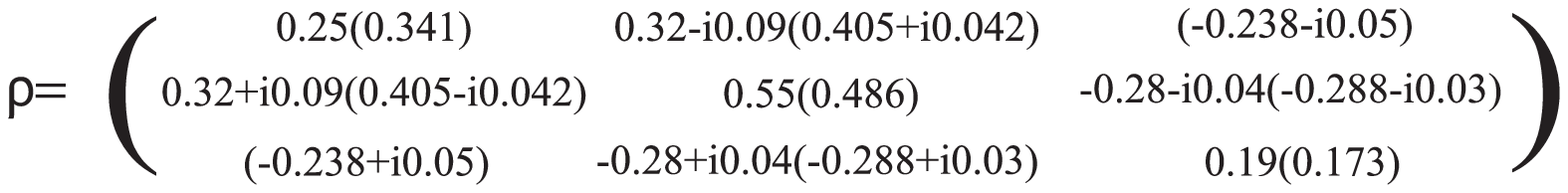}
  \caption{Density matrix that is measure for the input state given by the
angle of the setting wave plate $\alpha = 25^0$. Theoretically
calculated values are placed in brackets.}
\end{figure}

\end{document}